\documentclass[aps,prd,preprint,groupedaddress]{revtex4-1}

 \usepackage{graphicx} 

\begin{document}

\title{  Heavy meson dissociation in a plasma with magnetic fields   }



\author{Nelson R. F. Braga}\email{braga@if.ufrj.br}
\affiliation{Instituto de F\'{\i}sica,
Universidade Federal do Rio de Janeiro, Caixa Postal 68528, RJ
21941-972 -- Brazil}

\author{Luiz F.  Ferreira}
\email{luizfaulhaber@if.ufrj.br}
\affiliation{Instituto de F\'{\i}sica,
Universidade Federal do Rio de Janeiro, Caixa Postal 68528, RJ
21941-972 -- Brazil}


\begin{abstract}

The fraction of heavy vector mesons detected after a heavy ion collision provides information about the possible formation of a plasma state. 
An interesting framework for estimating the degree of dissociation of heavy mesons in a plasma 
is the holographic approach.  
It has been recently shown that a consistent picture for the thermal behavior of charmonium and bottomonium states  in a thermal medium emerges from holographic bottom up models.
A crucial ingredient in this new approach is the appropriate description of decay constants, since they are related to the heights of the quasiparticle peaks of the finite temperature spectral function.

Here we extend this new holographic model in order to study the effect of magnetic fields on the thermal spectrum of heavy mesons.  The motivation is that very large magnetic fields are  present in non central heavy ion collisions and  this  could imply a change in  the dissociation scenario. 
The thermal spectra of $ c  \bar c$ and $ b \bar b  \, $ S wave states is obtained for different temperatures and different values of the  magnetic $e B$ field.  
     
\end{abstract}


\maketitle

\section{ Introduction }   

A consistent picture for the thermal behavior of heavy vector mesons in a plasma was 
obtained recently using holographic bottom up models\cite{Braga:2016wkm,Braga:2017oqw,Braga:2017bml}.
 A central point in these works  is the connection between 
 the finite temperature spectral function and the zero temperature decay constants.
The spectral function -- that describes the thermal behavior of quasiparticles inside a thermal medium  --  is the imaginary part of the retarded Green's function.
At zero temperature,  the essential part of the Green's function has the following spectral decomposition in terms of masses $m_n$ and   decay constants $f_n$  of the states:  
 \begin{equation}
\Pi (p^2)  \sim  \sum_{n=1}^\infty \, \frac{f_n^ 2}{(- p^ 2) - m_n^ 2 + i \epsilon} \,.
\label{2point}
\end{equation} 
The imaginary part of this expression is  a sum of Dirac deltas with coefficients proportional to the square of the decay constants: $ f_n^2 \, \delta ( - p^2 - m_n^2 ) $. 
At finite temperature, the quasi-particle states appear in the spectral function as smeared -- finite size --  peaks with a height  that decrease as the temperature $T$ and/or the density $\mu$ of the medium increase. This analysis strongly suggests that in order to extend  a hadronic model  to finite temperature, the zero temperature case should provide  a  consistent description of decay constants.

  Decay constants for mesons are  associated with non-hadronic decay. They are proportional to the transition matrix from a state at excitation level $n$ to the hadronic vacuum:  $  \langle 0 \vert \, J_\mu (0)  \,  \vert n \rangle = \epsilon_\mu f_n m_n $.  Experimental data show that for heavy vector mesons the decay constants decrease monotonically with radial excitation level, as revised in \cite{Braga:2017oqw,Braga:2017bml}.
    
 Holographic models, inspired in the AdS/CFT 
 correspondence\cite{Maldacena:1997re,Gubser:1998bc,Witten:1998qj}, provide nice estimates for hadronic masses. However neither the hard wall  
 \cite{Polchinski:2001tt,BoschiFilho:2002ta,BoschiFilho:2002vd} , the soft wall \cite{Karch:2006pv} 
 or the D4-D8 \cite{Sakai:2004cn}  models provide decay constants decreasing with excitation level.
 
  An alternative bottom up holographic model was developed in ref.  \cite{Braga:2015jca} 
 in order to overcome this problem. The decay constants are obtained from two point correlators of gauge theory operators  calculated at a  finite value  of the radial coordinate of AdS space.   This way an extra energy parameter, associated with  an ultraviolet (UV)  energy scale, is introduced  in the model. 
The extension of  this model to finite temperature in \cite{Braga:2016wkm} and finite density in \cite{Braga:2017oqw} provided  consistent pictures for  the dissociation of heavy vector mesons is the plasma. 
An improved version of the model of ref. \cite{Braga:2015jca} ,  that provides a better fit for the charmonium states at zero temperature and thus a better picture  for the finite temperature and density cases,  was then proposed  in \cite{Braga:2017bml}.

An interesting tool to investigate the possible existence of a plasma state in a heavy ion collision is 
to analyze the fraction of heavy vector mesons produced. 
 The suppression of such particles indicates their dissociation in the medium\cite{Matsui:1986dk} (see also \cite{Satz:2005hx}).  This effect   
corresponds to a decrease in  the height of the quasi particle peaks of the 
spectral function.  The influence  of temperature and density of the medium in heavy vector meson spectral functions was studied in \cite{Braga:2016wkm,Braga:2017oqw,Braga:2017bml}.  However, there is another important factor that deserves consideration. 
In non central heavy ion collisions  strong magnetic fields can be produced for short time scales\cite{Kharzeev:2007jp,Fukushima:2008xe,Skokov:2009qp}. 

The presence of a magnetic field $e B$ has important consequences for hadronic matter. Lattice results \cite{Bali:2011qj} indicate a decrease in the QCD deconfinement temperature with increasing $ e B $ field. 
Similar results show up also from  the MIT bag model\cite{Fraga:2012fs} and also from  the holographic D4-D8 model\cite{Ballon-Bayona:2013cta}. The effect of a magnetic field in the transition temperature of a plasma has been studied using holographic models in many works, as for example \cite{Mamo:2015dea,Dudal:2015wfn,Evans:2016jzo,Li:2016gfn,Ballon-Bayona:2017dvv,Rodrigues:2017cha}.

Here we extend the holographic bottom up model of \cite{Braga:2017bml}  in order to include the presence of a magnetic field. This way it is possible to investigate the change in the spectral function peaks that represent the quasiparticle heavy meson states as a function of the intensity of the $e B$ field.  In section II we describe the model at zero temperature showing the results for masses and decay constants. Then, in section III we present the extension to finite temperature in the presence of a magnetic field. Section IV is devoted to show  how to calculate the spectral functions. Finally, in section V we present the results as discuss their implication in terms of heavy vector meson dissociation.  
 
 \section{Holographic Model }
 
The  model proposed in ref.\cite{Braga:2017bml}  was conceived for describing  charmonium states. At  zero temperature the background geometry is the standard 5D anti-de Sitter space-time 
\begin{equation}
 ds^2 \,\,= \,\, \frac{R^2}{z^2 }(-dt^2 + d\vec{x}\cdot d\vec{x} + dz^2)\,.
\end{equation}
The mesons are described by a vector  field $V_m = (V_\mu,V_z)\,$ ($\mu = 0,1,2,3$), which is dual to the gauge theory current $ J^\mu = \bar{\psi}\gamma^\mu \psi \,$. The action is:
\begin{equation}
I \,=\, \int d^4x dz \, \sqrt{-g} \,\, e^{- \phi (z)  } \, \left\{  - \frac{1}{4 g_5^2} F_{mn} F^{mn}
\,  \right\} \,\,, 
\label{vectorfieldaction}
\end{equation}
where $F_{mn} = \partial_m V_n - \partial_n V_m$ and $\phi(z)$ is a  background  dilaton field that here we choose to have the form
\begin{equation}
\phi(z)=k^2z^2+Mz+\tanh\left(\frac{1}{Mz}-\frac{k}{ \sqrt{\Gamma}}\right) \,,
\label{dilatonModi}
\end{equation} 
in order to represent both charmonium and bottomonium states. The parameter $k$ represents the  quark mass, $\Gamma $ the string tension of the strong quark anti-quark interaction and $M$ is a mass scale associated with non hadronic decay. 

 Choosing the gauge $V_z=0$  the equation of motion for the transverse (1,2,3) components  of the  field, denoted generically as $V$, in momentum space reads
\begin{equation}
\partial_{z} \left[ e^{-B(z)} \partial_{z} V \right]-p^2 e^{-B(z)}V=0, 
\label{eqmotion}
\end{equation}
where   $B(z)$ is  
\begin{equation}
B(z)=\log\left(\frac{z}{R}\right)+\phi(z)\,.
\label{B}
\end{equation}
Equation of motion (\ref{eqmotion}) presents a discrete spectrum of normalizable solutions, $ V(p,z)=\Psi_n(z)$ that satisfy the boundary conditions $ \Psi_n(z=0)=0$ for $p^2=-m_{n}^{2}$ where $m_n$ are the masses of the corresponding meson states. The eigenfunctions $\Psi_n(z)$ are  normalized according to:
\begin{equation}
\int^{\infty}_{0}dz \ e^{-B(z)} \ \Psi_n(z)\Psi_m(z)=\delta_{mn} \,.
\label{Normalization}
\end{equation}
Decay constants are  proportional to the transition matrix from  the vector meson $n$ excited state to the vacuum:  $  \langle 0 \vert \, J_\mu (0)  \,  \vert n \rangle = \epsilon_\mu f_n m_n $. 
They are calculated holographically  in the same way as in the soft wall model:
\begin{equation}
f_n=\frac{1}{g_{5} m_{n}}\lim\limits_{z \rightarrow 0} \left( e^{-B(z)}\Psi_n(z)\right) \,.
\label{decayconstant}
\end{equation}

 The  values of the parameters that  describe  charmonium and bottomonium are respectively:
\begin{equation}
  k_c = 1.2  \, {\rm GeV } ; \,\,   \sqrt{\Gamma_c } = 0.55  \, {\rm GeV } ; \,\, M_c=2.2  \, {\rm GeV }\,;
  \label{parameters1}
  \end{equation}   
\begin{equation}
  k_B = 2.45  \, {\rm GeV } ; \,\,   \sqrt{\Gamma_B } = 1.55  \, {\rm GeV } ; \,\, M_B=6.2  \, {\rm GeV }\,.
  \label{parameters2}
  \end{equation}  
 The procedure to calculate masses and decay constants is to find the normalizable solutions $ \Psi_n(z)$ of eq. (\ref{eqmotion}), with the background of eq. (\ref{dilatonModi}),  that vanish at $z = 0$.  Then the numerical solutions are used in 
 eq. (\ref{decayconstant}).  Tables \textbf{1} and \textbf{2} show the results for charmonium and bottomonium respectively. For comparison, the experimental data from ref. (\cite{Agashe:2014kda}) is show inside parenthesis.  Note that the decay constants decrease with  radial excitation level.

\begin{table}[h]
\centering
\begin{tabular}[c]{|c||c||c|}
\hline 
\multicolumn{3}{|c|}{  Holographic (and experimental)  Results for Charmonium   } \\
\hline
 State &  Mass (MeV)     &   Decay constants (MeV) \\
\hline
$\,\,\,\, 1S \,\,\,\,$ & $ 2943 \,\, (3096.916\pm 0.011)  $  & $ 399 \, (416 \pm 5.3)$ \\
\hline
$\,\,\,\, 2S \,\,\,\,$ & $  3959 \,\, (3686.109 \pm 0.012) $   & $ 255  \, (296.1 \pm 2.5)$  \\
\hline 
$\,\,\,\,3S \,\,\,\,$ & $  4757 \,\, (4039 \pm 1 ) $   & $198 \, ( 187.1  \pm 7.6) $ \\ 
\hline
$ \,\,\,\, 4S  \,\,\,\,$ & $ 5426\,\,  (4421 \pm 4)  $  & $ 169 \,  (160.8  \pm 9.7)$ \\
\hline
\end{tabular}   
\caption{Holographic masses and the corresponding decay constants for the Charmonium S-wave resonances. Experimental values inside parenthesis for comparison.  }
\end{table}

\begin{table}[h]
\centering
\begin{tabular}[c]{|c||c||c|}
\hline 
\multicolumn{3}{|c|}{  Holographic (and experimental) Results for Bottomonium   } \\
\hline
 State &  Mass (MeV)     &   Decay constants (MeV) \\
\hline
$\,\,\,\, 1S \,\,\,\,$ & $ 6905 \,\,(9460.3\pm 0.26) $  & $ 719 \,( 715.0 \pm 2.4) $ \\
\hline
$\,\,\,\, 2S \,\,\,\,$ & $   8871 \,( 10023.26 \pm 0.32) $   & $ 521 \,(497.4 \pm 2.2) $  \\
\hline 
$\,\,\,\,3S \,\,\,\,$ & $  10442 \, \,( 10355.2 \pm 0.5) $   & $427 \, (430.1  \pm 1.9) $ \\ 
\hline
$ \,\,\,\, 4S  \,\,\,\,$ & $ 11772 \, (10579.4 \pm 1.2)  $  & $ 375 \,(340.7  \pm 9.1)$ \\
\hline
\end{tabular}   
\caption{Holographic masses and the corresponding decay constants for the Bottomonium S-wave resonances. Experimental values inside parenthesis for comparison. }
\end{table}

\section{ Plasma with magnetic field  } 
 Let us now extend the model to finite temperature and in the presence of magnetic field, assumed for simplicity to be constant in time and homogeneous in space.  
 The extension to finite temperature  is obtained replacing AdS space by  a Schwarzschild AdS black hole.  The presence of a magnetic field in the gauge theory side of gauge/gravity duality can also be represented geometrically in the gravity 
 side\cite{DHoker:2009mmn,DHoker:2009ixq}.  
 The Einstein-Maxwell action is given by:
\begin{equation}\label{bulkaction}
S=\frac{1}{16 \pi G_{5}}\int d^{5}x \sqrt{-g}\left(R-F^{MN}F_{MN}+\frac{12}{L^2} \right)+S_{GH}
\end{equation}
with $F_{MN}$ is the electromagnetic field strength, $R$ is the Ricci scalar and $\Lambda$=$\, - \frac{12}{L^2}$ is the negative cosmological constant. The second term in eq. (\ref{bulkaction}) is the  Gibbons-Hawking surface term.  
  
The equations of motion obtained from eq.(\ref{bulkaction}) are
\begin{eqnarray}
R_{MN}=&-&\frac{4}{L^2}g_{MN}-\frac{g_{MN}}{3}F^{PQ}F_{PQ}
+ 2F_{MP}F_{\,\,\,N}^{P},
\end{eqnarray}
\begin{eqnarray}
\nabla_{M}F_{MN}=0.
\end{eqnarray}
  In order to represent a magnetic field $eB$ we will use a  Black hole solution  studied in  \cite{Dudal:2015wfn}
\begin{equation}
 ds^2 \,\,= \,\, \frac{R^2}{z^2}  \,  \Big(  -  f(z) dt^2 + \frac{dz^2}{f(z) }  +( dx_1^2+dx_2^2)d(z)+dx_3^2q(z)  \Big)   \,.
 \label{metric2}
\end{equation}
In this expression the factors  are
\begin{equation}
f (z) = 1 - \frac{z^ 4}{z_h^4}+\frac{2}{3}\frac{e^2B^2z^4}{1.6^2}\ln\left(\frac{z}{z_h}\right)\, ,
\end{equation}
\begin{equation}
q(z) = 1 + \frac{8}{3}\frac{e^2B^2}{1.6^2}\int^{1/z}_{+\infty}dx\frac{\ln{(z_hz)}}{z^3(z^2-\frac{1}{z_h^4x^2})}.
\end{equation}
\begin{equation}
d(z) = 1 - \frac{4}{3}\frac{e^2B^2}{1.6^2}\int^{1/z}_{+\infty}dx\frac{\ln{(z_hz)}}{z^3(z^2-\frac{1}{z_h^4x^2})}.
\end{equation}
where  $z_h$  is the horizon position and $ eB$ is the boundary magnetic field that is in the $x_3 $ direction. 
The temperature of the black hole and of the gauge theory is: 
\begin{equation} 
T =  \frac{\vert  f'(z)\vert_{(z=z_h)}}{4 \pi  } = \frac{1}{4\pi}\left\vert \frac{4}{z_h}-\frac{2}{3}e^2B^2z_{h}^3\right\vert \,.
\label{temp}
\end{equation}
 
 We assume that the geometry is not modified by the presence of the dilaton background. So, the action has the same form  of eq. (\ref{vectorfieldaction}) with the dilaton background $ \phi (z) $ of 
eq. (\ref{dilatonModi})  but with metric the (\ref{metric2}).
Now the equations of motion have to be solved numerically. In the next section we discuss how to solve them with the appropriate boundary conditions using the membrane paradigm.

\section{Spectral Function}
  
The spectral functions for heavy vector mesons will be calculated using the membrane paradigm \cite{Iqbal:2008by} (see also \cite{Finazzo:2015tta}). Let us see how this formalism works for a vector field $V_\mu$ in the bulk, dual to the electric  current operator $J_{\mu}$. We consider a general black brane background  of the form
\begin{equation}
\label{metric3}
ds^{2}=-g_{tt}dt^2+g_{zz}dz^{2}+g_{x_1 x_1}dx^{2}_{1}+g_{x_2 x_2}dx^{2}_{2}+g_{x_3 x_3}dx^{2}_{3}\,,
\end{equation}
where we assume the boundary is at a position  $z=0$ and  the horizon  position $z = z_{h}$ is given by the root of $g_{tt}(z_{h})=0$. The bulk action for the vector field  is
\begin{equation}\label{Sigma2}
S=-\int d^{5}x \, \sqrt{-g}\frac{1}{4 g_5^2 \, h(z)}F^{m n}F_{m n}\,,
\end{equation}
where $h(z)$ is  a z-dependent coupling. The equation of motion  that follows from this action is:
\begin{equation}\label{eqms}
\partial^{m} \left( \frac{\sqrt{-g}}{h(z)}  F_{m n } \right)=0 \,.
\end{equation}
The conjugate momentum to the gauge field with respect to a foliation by constant z-slices is given by:
\begin{equation}\label{momentum}
j^{\mu}=-\frac{1}{h(z)}\sqrt{-g}F^{z \mu} \,.
\end{equation}

We assume that the metric (\ref{metric3}) satisfies $ g_{x_1 x_1} =  g_{x_2 x_2} $ that includes our case of interest:  metric (\ref{metric2}) that represents a magnetic field in the $x_3$ direction.
We also assume solutions for the vector field that do not depend on the coordinates  $x_1$ and $x_2$, like a plane wave with spatial momentum in the $x_3$ direction. 
Equations of motion  (\ref{eqms})  can be separated in  two different channels: a longitudinal one  involving fluctuations along $(t, x_3)$ and a transverse channel involving fluctuations along  spatial directions  $(x_1,x_2)$. 
Using eq. (\ref{momentum}),  the components $t$, $x_3$ and $z$ of  eq. (\ref{eqms}) can be written respectively as
\begin{equation}\label{Maxwell1}
-\partial_{z}j^{t}-\frac{\sqrt{-g}}{h}g^{tt}g^{x_3 x_3}\partial_{x_3}F_{x_3t}=0\,,
\end{equation}
\begin{equation}\label{Maxwell2}
-\partial_{z}j^{x_3}+\frac{\sqrt{-g}}{h}g^{tt}g^{x_3 x_3}\partial_{t}F_{x_3t}=0\,,
\end{equation}
\begin{equation}\label{Maxwellz}
\partial_{x_3}j^{x_3}+\partial_t j^{t}=0\,.
\end{equation}
 From the Bianchi identity one finds the relation:
\begin{equation}\label{Bianchi}
\partial_{z}F_{x_3t}-\frac{h(z)}{\sqrt{-g}}g_{zz}g_{x_3 x_3}\partial_{t}j^{z}-\frac{h(z)}{\sqrt{-g}}g_{tt}g_{x_3 x_3}\partial_{x_3}j^{t}=0\,.
\end{equation} 
Now, one can define a z-dependent "conductivity" for the longitudinal channel given by:
\begin{equation}\label{longitudinal}
\bar{\sigma}_{L}(\omega,\vec{p},z)=\frac{j^{x_3}(\omega,\vec{p},z)}{F_{x_3 t}(\omega,\vec{p},z)}\,.
\end{equation} 
Taking a derivative of eq. (\ref{longitudinal}) one finds:   
\begin{equation}\label{dlongitudinal}
\partial_z \bar{\sigma}_{L}=\frac{\partial_z j^{x_3}}{F_{x_3t}}-\frac{ j^{x_3}}{F_{x_3 t}^2}\partial_z F_{x_3 t}\,.
\end{equation} 
Now using  eqs.  (\ref{Maxwell1}),  (\ref{Maxwellz}) and (\ref{Bianchi}) and considering a plane wave solution with momentum  $p=(\omega,0,0,p_3)$, the previous equation for $\bar{\sigma}_{L}$ becomes 
\begin{equation}\label{Membrane}
\partial_{z}\bar{\sigma}_{L}=-i\omega\sqrt{\frac{g_{zz}}{g_{tt}}}\left[ \Sigma_{L}(z)-\frac{\bar{\sigma}_{L}^{2}}{\Sigma_{L}(z)} \left( 1-\frac{p_3^2}{\omega^2} \frac{g^{x_3 x_3}}{g^{tt}}          \right)    \right] \,,
\end{equation}
where 
\begin{equation}\label{Sigma}
\Sigma_{L} (z)=\frac{1}{h(z)}\sqrt{\frac{-g}{g_{zz}g_{tt}}}g^{x_3 x_3} \,.
\end{equation}
Similarly, the transverse channel is governed by a dynamical equation
\begin{equation}\label{transverse}
-\partial_z j^{x_1}-\frac{\sqrt{-g}}{h(z)}g^{tt}g^{x_1 x_{1}}\partial_{t}F_{tx_1}+\frac{\sqrt{-g}}{h(z)}g^{x_3 x_3}g^{x_1 x_1}\partial_{x_{3}}F_{x_{3}x_{1}}=0\,.
\end{equation}
and two constraints from the Bianchi identity
\begin{eqnarray}\label{trans2verse}
\partial_{z}F_{x_1 t}-\frac{h(z)}{\sqrt{-g}} g_{zz}g_{x_1 x_1} \partial_{t}j^{x_1}&=&0 \,,\nonumber\\
\partial_{x_{3}}F_{t x_{1}}+\partial_{t}F_{x_{1} x_{3}}&=&0\,.
\end{eqnarray}
One can define a z-dependent "conductivity"  also for the transverse channel: 
\begin{equation}\label{sigmatransverse}
\bar{\sigma}_{T}(\omega,\vec{p},z)=\frac{j^{x_1}(\omega,\vec{p},z)}{F_{x_1 t}(\omega,\vec{p},z)}\,.
\end{equation}
Following the same procedure used for the longitudinal channel, using eqs. (\ref{transverse}-\ref{trans2verse}) one finds the equation for $\bar{\sigma}_{T}$:
\begin{equation}\label{Membrane2}
\partial_{z}\bar{\sigma}_{T}=i\omega\sqrt{\frac{g_{zz}}{g_{tt}}}\left[ \frac{\bar{\sigma}_{T}^{2}}{\Sigma_{T}(z)} - \Sigma_{T}(z)\left( 1-\frac{p_3^2}{\omega^2} \frac{g^{x_3 x_3}}{g^{tt}}          \right)    \right] \,.
\end{equation}
where
\begin{equation}\label{Sigma}
\Sigma_{T} (z)=\frac{1}{h(z)}\sqrt{\frac{-g}{g_{zz}g_{tt}}}g^{x_1 x_1} \,.
\end{equation}
In the zero  momentum limit, the equations  (\ref{Membrane}) and (\ref{Membrane2}) became respectively:
\begin{equation}\label{Membranep0}
\partial_{z}\bar{\sigma}_{L}=i\omega\sqrt{\frac{g_{zz}}{g_{tt}}}\left[\frac{\bar{\sigma}_{L}^{2}}{\Sigma_{L}(z)}  -\Sigma_{L}(z)   \right] \,,
\end{equation}
\begin{equation}\label{Membrane2p0}
\partial_{z}\bar{\sigma}_{T}=i\omega\sqrt{\frac{g_{zz}}{g_{tt}}}\left[ \frac{\bar{\sigma}_{T}^{2}}{\Sigma_{T}(z)} - \Sigma_{T}(z) \right] \,.
\end{equation}
Using the Kubo's formula it is possible relate the $ z=0$ values of the five dimensional ``conductivity'' to the retarded Green's function:
\begin{equation}\label{AC1}
\sigma_{L}(\omega) =-\frac{G_{R}^{L}(\omega)}{i \omega}\equiv \bar{\sigma}_{L}(\omega,0)\,.
\end{equation}
\begin{equation}\label{AC2}
\sigma_{T}(\omega) =-\frac{G_{R}^{T}(\omega)}{i \omega}\equiv \bar{\sigma}_{T}(\omega,0)\,.
\end{equation}
where $\sigma_{T}$ is  the AC  conductivity in transverse channel and $\sigma_{L}$ is the AC conductivity in longitudinal channel.

\begin{figure}[h]
\label{g67}
\begin{center}
\includegraphics[scale=0.45]{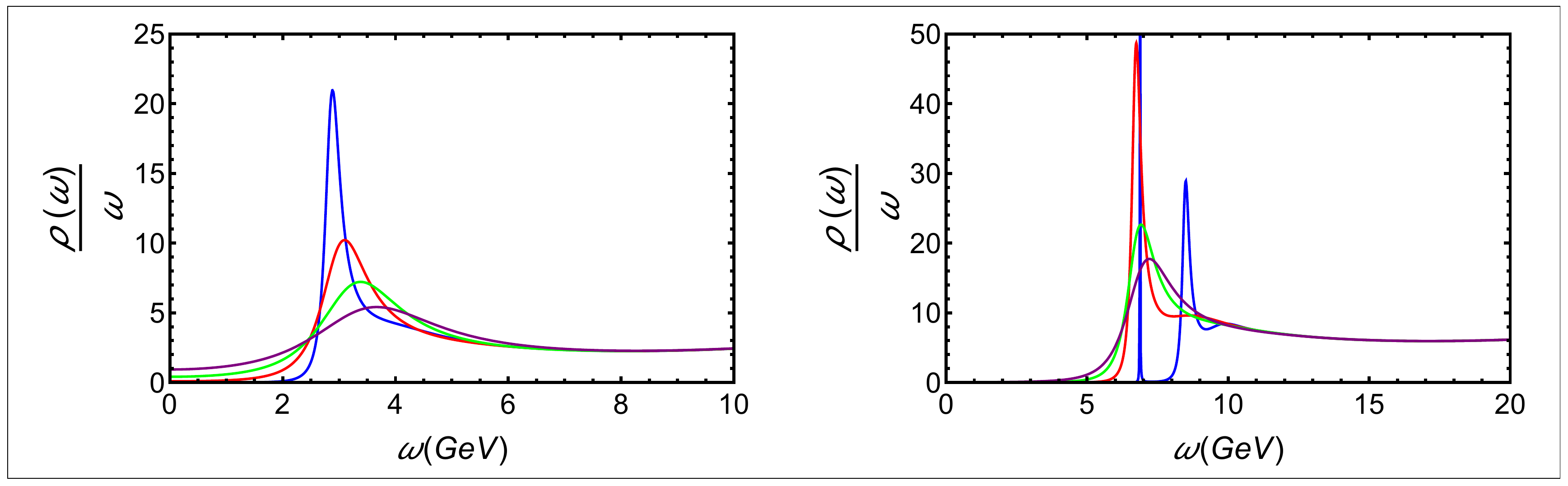}
\end{center}
\caption{Spectral functions for charmonium (left panel) and bottomonium (right panel) at 4 representative  values of the temperature: blue line T=195 MeV ; red line T =330 MeV; green line T = 465 MeV;  purple line T = 600 MeV.}
\end{figure}

\begin{figure}[h]
\label{g67}
\begin{center}
\includegraphics[scale=0.35]{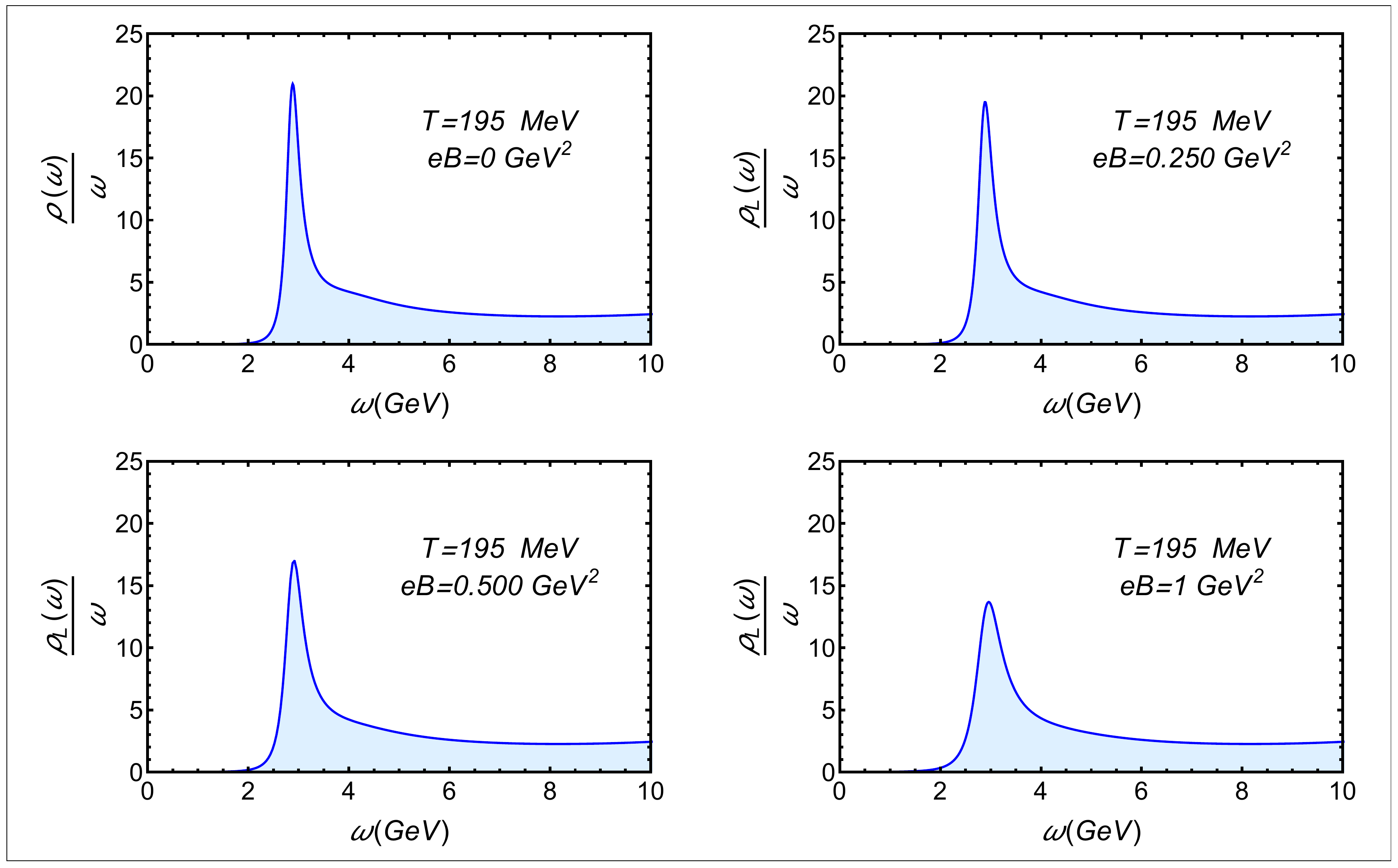}
\end{center}
\caption{Spectral functions for charmonium at T= 195  MeV  for different values of a magnetic field  parallel to the polarization}
\end{figure}

\begin{figure}[h]
\label{g67}
\begin{center}
\includegraphics[scale=0.35]{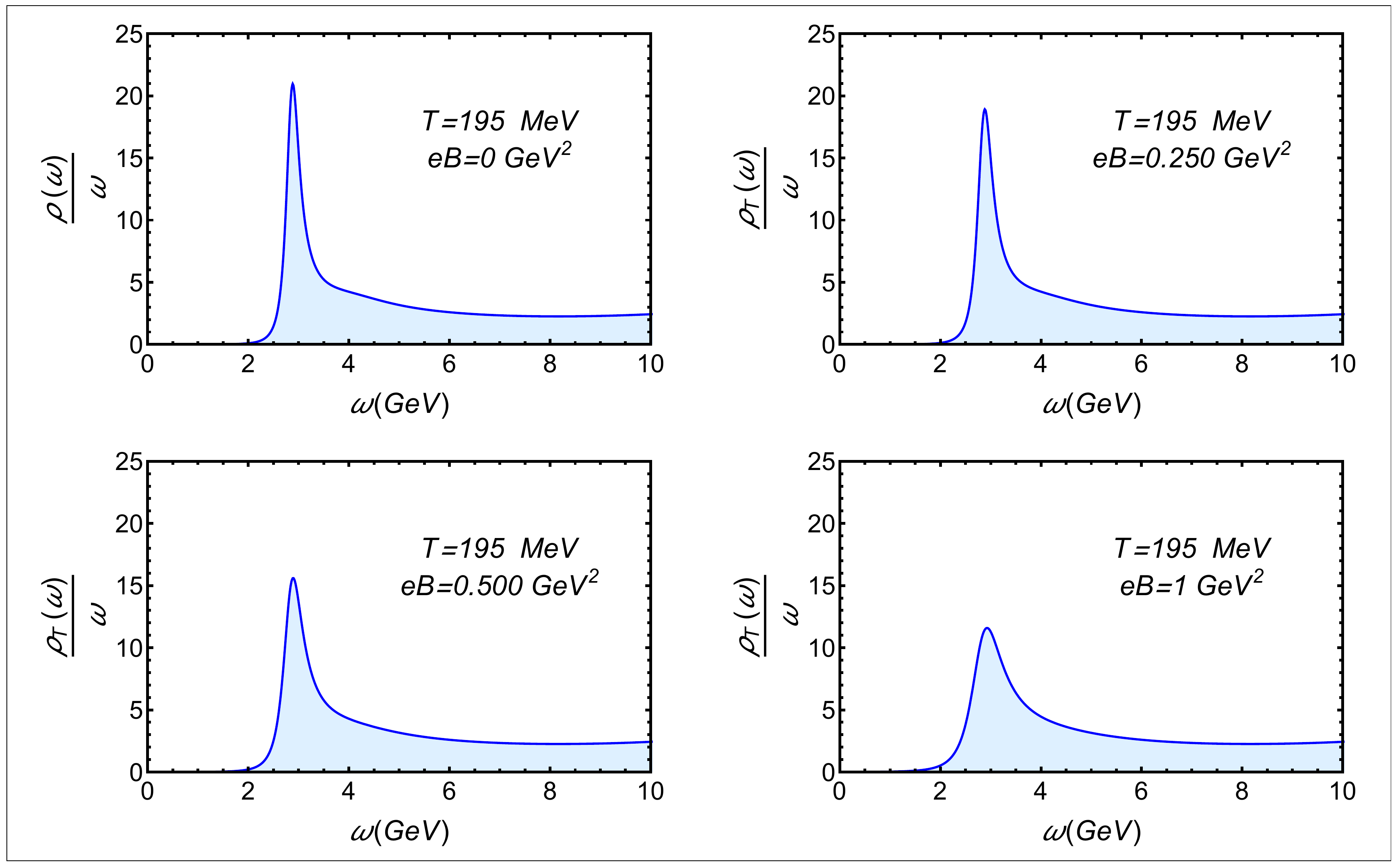}
\end{center}
\caption{ Spectral functions for charmonium at T= 195  MeV  for different values of a magnetic field  perpendicular to the polarization}
\end{figure}

In order to apply the membrane paradigm to the model of the previous section, we use the metric (\ref{metric2}) and $ h(z) = e^{\phi (z)  } $, with $\phi (z)$ defined in eq. (\ref{dilatonModi}), 
 in the flow equations  (\ref{Membranep0}) and (\ref{Membrane2p0}). The anisotropic metric (\ref{metric2})  leads to two different conductivities \cite{Rebhan:2011vd}:
\begin{equation}\label{MembraneF2L}
\partial_{z}\bar{\sigma}_{L}(\omega,z)= \frac{i\omega}{f(z)\bar{\Sigma}_{L}(z)} \left[ \bar{\sigma}_{L}(\omega,z)^{2}-\bar{\Sigma}_{L}(z)^{2}  \right] \,,
\end{equation}      
with $\bar{\Sigma}_{L}(z)=\frac{d(z)}{\sqrt{q(z)}}e^{-\phi(z) }/z$  and 
\begin{equation}\label{MembraneF2T}
\partial_{z}\bar{\sigma}_{T}(\omega,z)= \frac{i\omega}{f(z)\bar{\Sigma}_{T}(z)} \left[ \bar{\sigma}_{T}(\omega,z)^{2}-\bar{\Sigma}_{T}(z)^{2}  \right] \,,
\end{equation} 
with $\bar{\Sigma}_{T}(z)=\sqrt{q(z)}e^{-\phi(z)}/z$ . The equations can be solved  requiring regularity at the horizon, one obtains the following conditions:
\begin{equation}\label{boundarycondition12}
\bar{\sigma}_{L}(\omega,z_h)=\bar{\Sigma}_{L}(z_{h})\,\,;\,\,\,
 \bar{\sigma}_{T}(\omega,z_h)=\bar{\Sigma}_{T}(z_{h})\,.
\end{equation}
The spectral function is obtained from the relation:   
\begin{equation}\label{spectralfunction}
\rho(\omega)\equiv -Im \ G_{R}(\omega)= \omega  Re \  \bar{\sigma}(\omega,0)\,.
\end{equation}
Note that when the magnetic field is zero, the metric (\ref{metric2}) became isotropic and the  both flow equations (\ref{MembraneF2T}) and (\ref{MembraneF2L})  have the same form:

\begin{equation}\label{MembraneISO}
\partial_{z}\bar{\sigma}(\omega,z)= \frac{i\omega}{f(z)\bar{\Sigma}(z)} \left[ \bar{\sigma}(\omega,z)^{2}-\bar{\Sigma}(z)^{2}  \right] \,,
\end{equation}
where $\bar{\sigma}=\bar{\sigma}_{T}=\bar{\sigma}_{L}$.

\begin{figure}[t]
\label{g67}
\begin{center}
\includegraphics[scale=0.35]{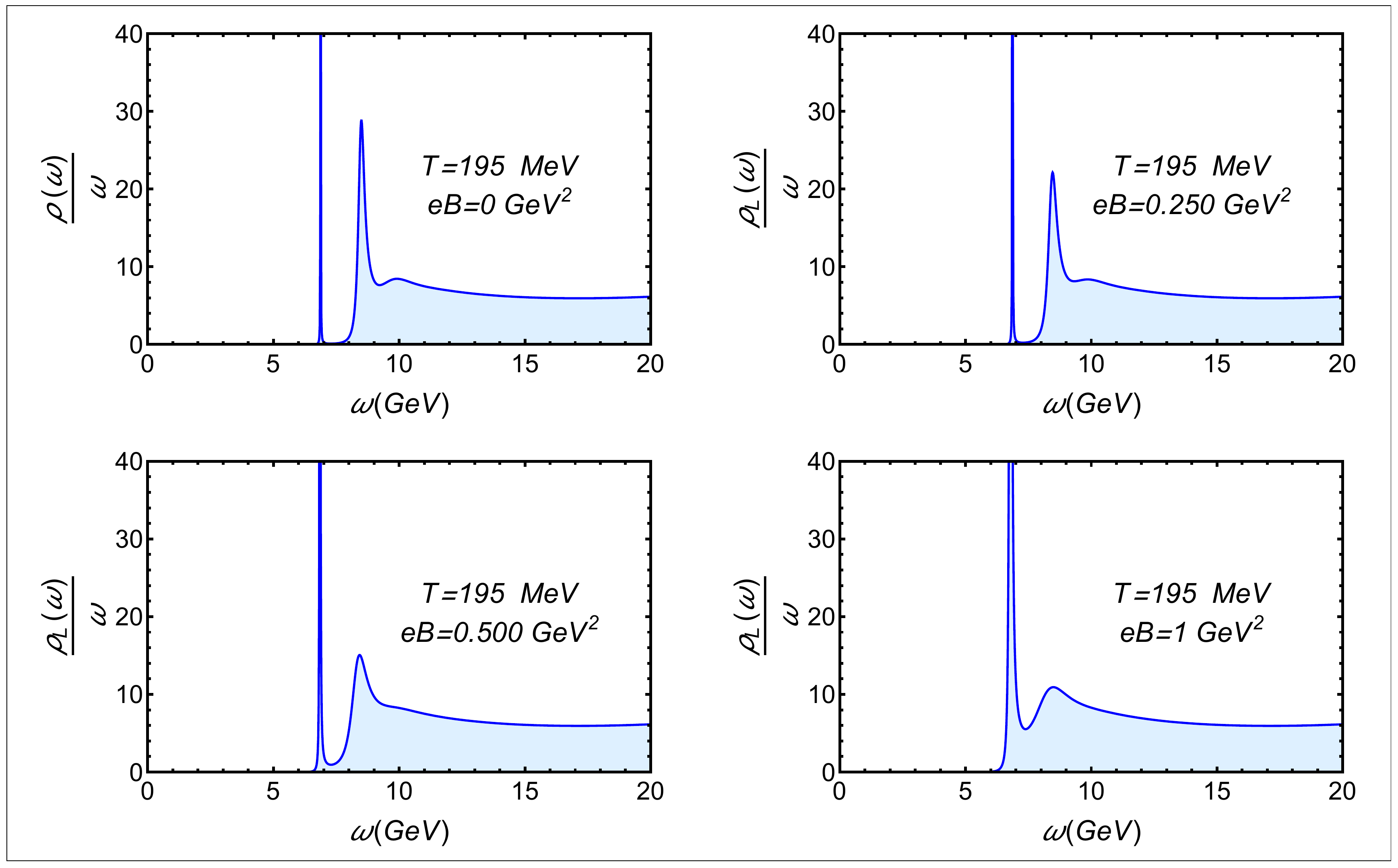}
\end{center}
\caption{Spectral functions for bottomonium at T= 195  MeV  for different values of a magnetic field  parallel to the polarization}
\end{figure}
 
\begin{figure}[t]
\label{g67}
\begin{center}
\includegraphics[scale=0.35]{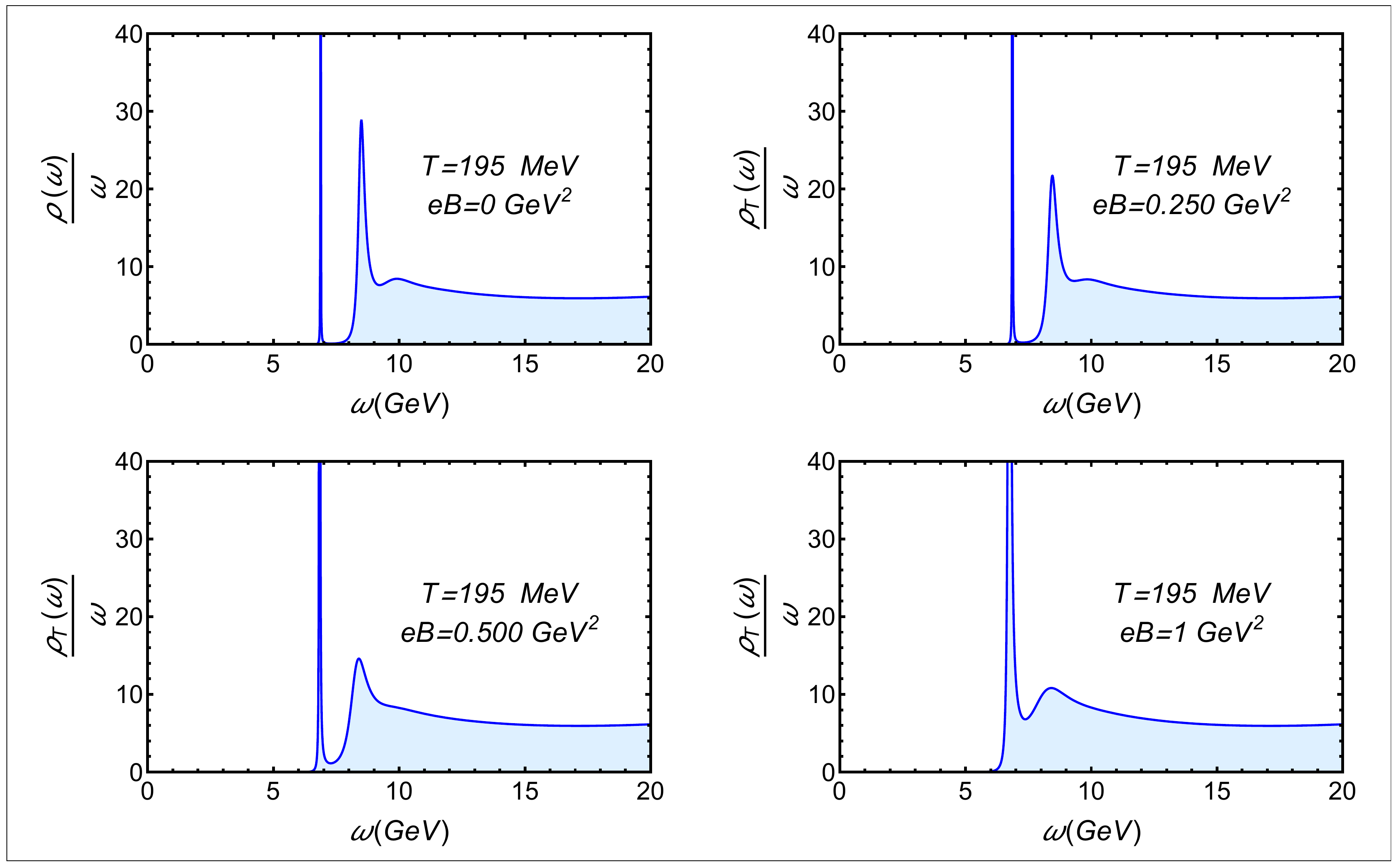}
\end{center}
\caption{ Spectral functions for bottomonium at T= 195  MeV  for different values of a magnetic field  perpendicular to the polarization}
\end{figure}

\begin{figure}[h]
\label{g67}
\begin{center}
\includegraphics[scale=0.35]{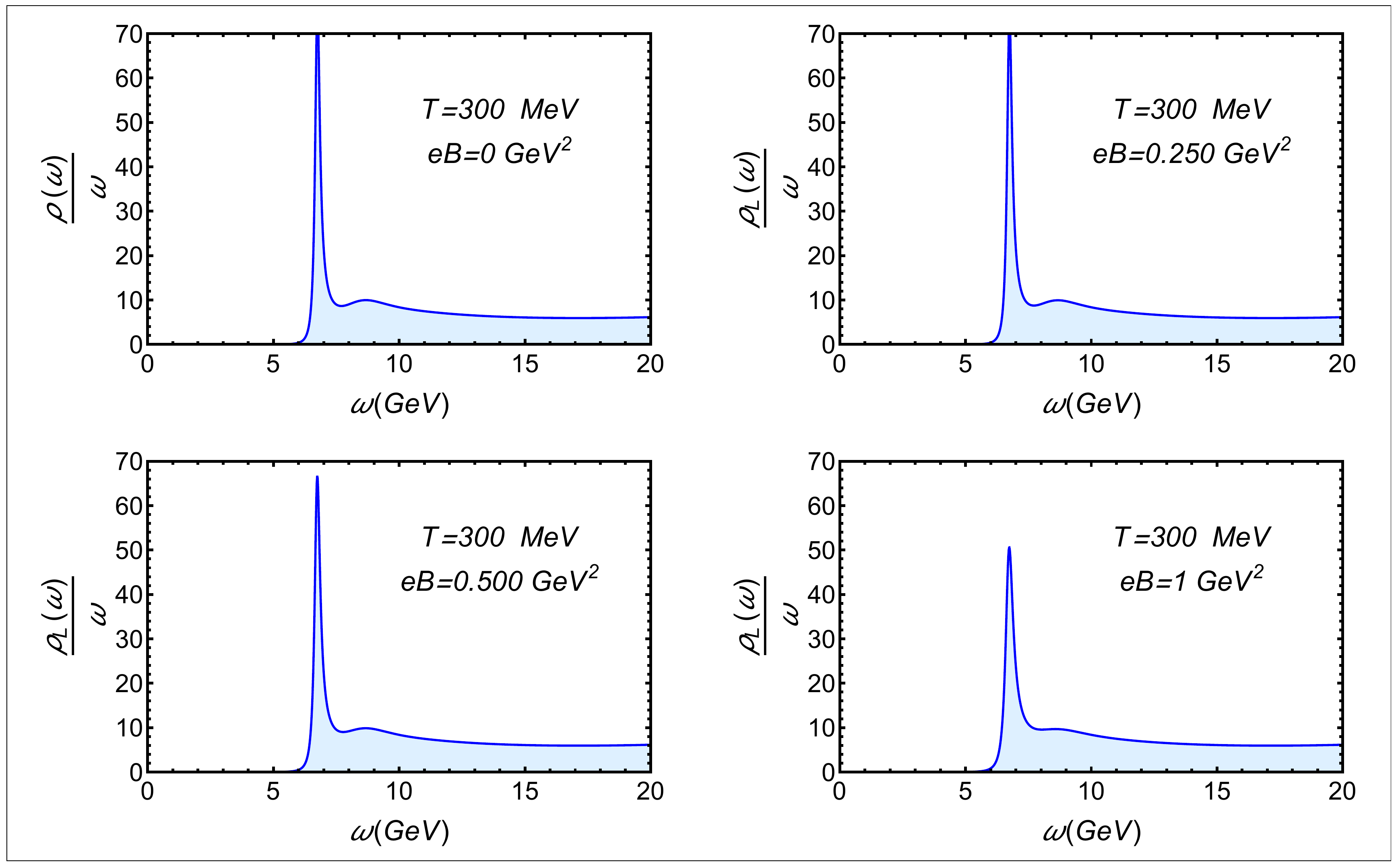}
\end{center}
\caption{Spectral functions for bottomonium at T= 300  MeV  for different values of a magnetic field  parallel to the polarization}
\end{figure}

\begin{figure}[h]
\label{g67}
\begin{center}
\includegraphics[scale=0.35]{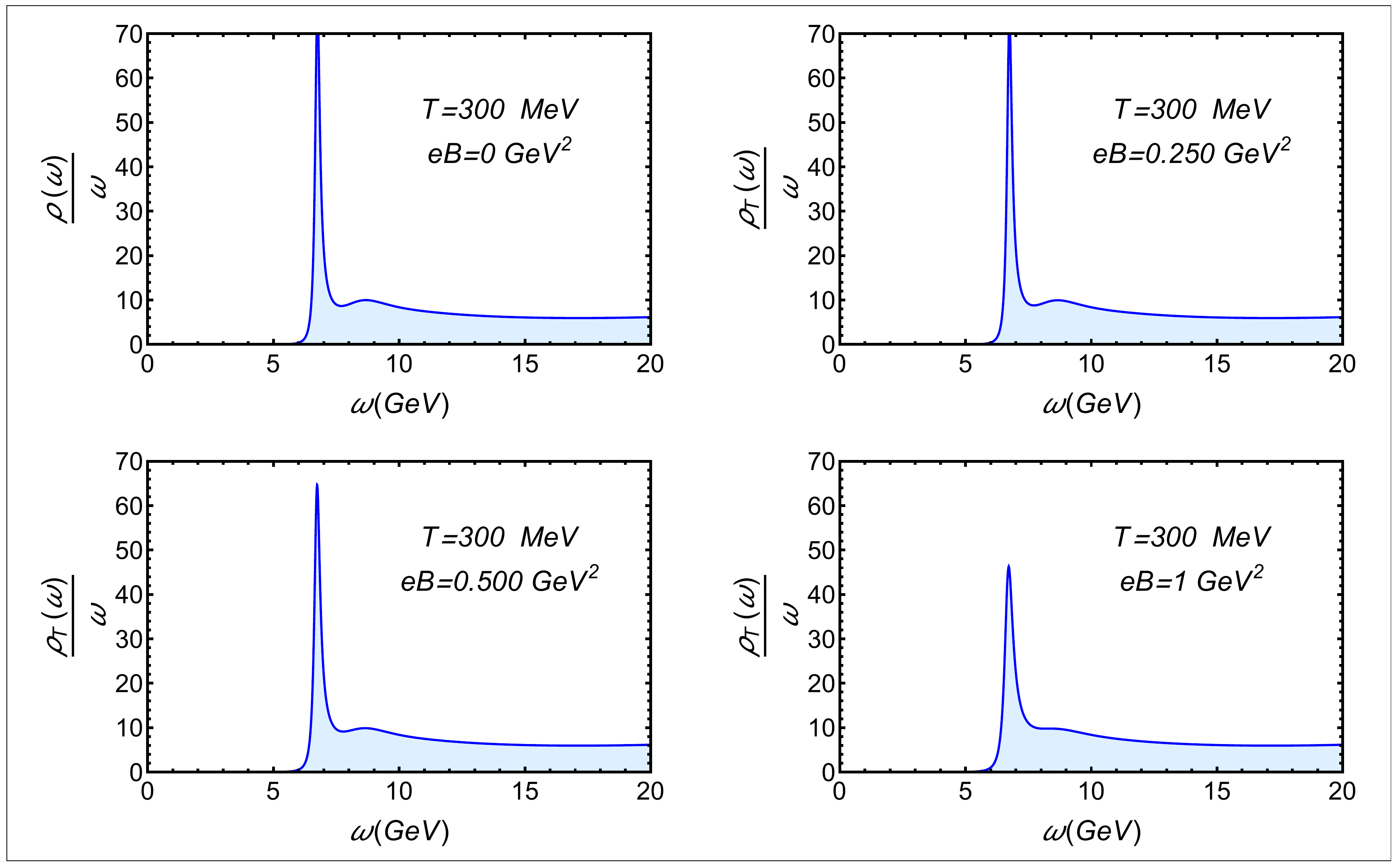}
\end{center}
\caption{ Spectral functions for bottomonium at T= 300  MeV  for different values of a magnetic field  perpendicular to the polarization}
\end{figure}

\section{Results and Discussion}   

The spectral functions for heavy vector mesons with polarization parallel (perpendicular) to the magnetic field are obtained evaluating  numerically equations (\ref{MembraneF2L})  (respectively  (\ref{MembraneF2T})), with the boundary conditions
(\ref{boundarycondition12})  described in the previous section. The parameters  used are the ones that provide the best fit in the  zero temperature case of section II namely those of eq. (\ref{parameters1}) for charmonium and those of eq. (\ref{parameters2}) for bottomonium. 

 In order to have a clear picture of the thermal behavior of the spectra when there is no magnetic field, we show in figure {\bf 1} the charmonium and bottomonium spectra at four representative temperatures.  One notes that for charmonium (left panel)  there is a peak at  $T= 195 $ MeV corresponding to the $ 1 S$  state, the $J /\Psi$. At higher temperatures the peak decreases  showing the dissociation  process.  For bottomonium there are two peaks at  $T= 195 $ MeV.  As the temperature increases the second state is strongly dissociated while the $ 1 S $ state has a much smaller dissociation effect, compared with charmonium. 
 
 Then we show in figures {\bf 2} and {\bf 3} the cases of charmonium at a temperature of 195 MeV for magnetic fields parallel and perpendicular, respectively, to the polarization.  This figures show that the dissociation effect increases with the magnetic field in both cases.  When the magnetic field is perpendicular  to the polarization of the meson the decrease in the spectral function peak is more noticeable then in the parallel case.

 Figures {\bf 4} and {\bf 5} show the cases of bottomonium  at  $T= 195 $ MeV for magnetic fields parallel and perpendicular, respectively, to the polarization. One notes the clear dissociation of the second state as the effect of increasing magnetic fields.  
The cases of bottomonium at  $T= 300 $ MeV are then presented in figures   {\bf 6} and {\bf 7}. 
The height of the peak of the  first state, the $\Upsilon $, decreases as the magnetic field increases. 
As it happens with charmonium the dissociation effect produced by the magnetic field is stronger when it is perpendicular to the polarization direction. 

The results provided by the model when there is no magnetic field are consistent with 
  results for quarkonium dissociation temperatures using lattice QCD, potential models and other studies as presented in \cite{Adare:2014hje}.  For the case with magnetic field, there is a study, only for  charmonium,  using a holographic model in ref. \cite{Dudal:2014jfa}. 
   The main result found there is that for magnetic fields perpendicular to the polarization the dissociation is stronger than without the field, while for $eB$ fields parallel to the polarization the dissociation is weaker than without the field.
  Our result is different. In the presence of  magnetic fields heavy vector mesons dissociate faster than without the field.
  The dissociation being faster for fields perpendicular to the polarization than for parallel fields.

 It is important to stress the fact  that our model, in contrast to ref.  \cite{Dudal:2014jfa}, reproduces the  behavior of the decay constants obtained experimentally.  As explained  in the introduction, this is very important in order to find a consistent description of the thermal behavior.

 \newpage

\noindent {\bf Acknowledgments:} N.B. is partially supported by CNPq (Brazil) under Grant No. 307641/2015-5 and L.F. is supported by CAPES (Brazil)


\begin{thebibliography}{ABC}

\bibitem{Braga:2016wkm} 
  N.~R.~F.~Braga, M.~A.~Martin Contreras and S.~Diles,
  Eur.\ Phys.\ J.\ C {\bf 76}, no. 11, 598 (2016)
  doi:10.1140/epjc/s10052-016-4447-4
  [arXiv:1604.08296 [hep-ph]].
  
\bibitem{Braga:2017oqw} 
  N.~R.~F.~Braga and L.~F.~Ferreira,
  Phys.\ Lett.\ B {\bf 773}, 313 (2017)
  doi:10.1016/j.physletb.2017.08.037
  [arXiv:1704.05038 [hep-ph]].
  
\bibitem{Braga:2017bml} 
  N.~R.~F.~Braga, L.~F.~Ferreira and A.~Vega,
  Phys.\ Lett.\ B {\bf 774}, 476 (2017)
  doi:10.1016/j.physletb.2017.10.013
  [arXiv:1709.05326 [hep-ph]]. 

\bibitem{Maldacena:1997re}
  J.~M.~Maldacena,
  Adv.\ Theor.\ Math.\ Phys.\  {\bf 2}, 231 (1998)
  [Int.\ J.\ Theor.\ Phys.\  {\bf 38}, 1113 (1999)].
 [arXiv:hep-th/9711200].

\bibitem{Gubser:1998bc}
  S.~S.~Gubser, I.~R.~Klebanov and A.~M.~Polyakov,
  Phys.\ Lett.\  B {\bf 428}, 105 (1998).
  [arXiv:hep-th/9802109].

\bibitem{Witten:1998qj}
  E.~Witten,
  Adv.\ Theor.\ Math.\ Phys.\  {\bf 2}, 253 (1998).
  [arXiv:hep-th/9802150].
    
\bibitem{Polchinski:2001tt}
  J.~Polchinski and M.~J.~Strassler,
  Phys.\ Rev.\ Lett.\  {\bf 88}, 031601 (2002)
  [arXiv:hep-th/0109174].

\bibitem{BoschiFilho:2002ta}
  H.~Boschi-Filho and N.~R.~F.~Braga,
  Eur.\ Phys.\ J.\  C {\bf 32}, 529 (2004)
  [arXiv:hep-th/0209080].
  
\bibitem{BoschiFilho:2002vd}
  H.~Boschi-Filho and N.~R.~F.~Braga,
  JHEP {\bf 0305}, 009 (2003)
  [arXiv:hep-th/0212207].
  

\bibitem{Karch:2006pv} 
  A.~Karch, E.~Katz, D.~T.~Son and M.~A.~Stephanov,
  Phys.\ Rev.\ D {\bf 74}, 015005 (2006)
  doi:10.1103/PhysRevD.74.015005
  [hep-ph/0602229].

\bibitem{Sakai:2004cn} 
  T.~Sakai and S.~Sugimoto,
  Prog.\ Theor.\ Phys.\  {\bf 113}, 843 (2005)
  doi:10.1143/PTP.113.843
  [hep-th/0412141].
   
   

\bibitem{Braga:2015jca} 
  N.~R.~F.~Braga, M.~A.~Martin Contreras and S.~Diles,
  Phys.\ Lett.\ B {\bf 763}, 203 (2016)
  doi:10.1016/j.physletb.2016.10.046
  [arXiv:1507.04708 [hep-th]].
  

\bibitem{Matsui:1986dk} 
  T.~Matsui and H.~Satz,
  Phys.\ Lett.\ B {\bf 178}, 416 (1986).
 doi:10.1016/0370-2693(86)91404-8.
 
 \bibitem{Satz:2005hx} 
  H.~Satz,
  J.\ Phys.\ G {\bf 32}, R25 (2006)
  doi:10.1088/0954-3899/32/3/R01
  [hep-ph/0512217].
 
 
    
    
\bibitem{Kharzeev:2007jp} 
  D.~E.~Kharzeev, L.~D.~McLerran and H.~J.~Warringa,
  Nucl.\ Phys.\ A {\bf 803}, 227 (2008)
  doi:10.1016/j.nuclphysa.2008.02.298
  [arXiv:0711.0950 [hep-ph]].
  
\bibitem{Fukushima:2008xe} 
  K.~Fukushima, D.~E.~Kharzeev and H.~J.~Warringa,
  Phys.\ Rev.\ D {\bf 78}, 074033 (2008)
  doi:10.1103/PhysRevD.78.074033
  [arXiv:0808.3382 [hep-ph]].
    
\bibitem{Skokov:2009qp} 
  V.~Skokov, A.~Y.~Illarionov and V.~Toneev,
  Int.\ J.\ Mod.\ Phys.\ A {\bf 24}, 5925 (2009)
  doi:10.1142/S0217751X09047570
  [arXiv:0907.1396 [nucl-th]].
    
    
\bibitem{Bali:2011qj} 
  G.~S.~Bali, F.~Bruckmann, G.~Endrodi, Z.~Fodor, S.~D.~Katz, S.~Krieg, A.~Schafer and K.~K.~Szabo,
  JHEP {\bf 1202}, 044 (2012)
  doi:10.1007/JHEP02(2012)044
  [arXiv:1111.4956 [hep-lat]].
    
    
\bibitem{Fraga:2012fs} 
  E.~S.~Fraga and L.~F.~Palhares,
  Phys.\ Rev.\ D {\bf 86}, 016008 (2012)
  doi:10.1103/PhysRevD.86.016008
  [arXiv:1201.5881 [hep-ph]].
    
    
\bibitem{Ballon-Bayona:2013cta} 
  A.~Ballon-Bayona,
  JHEP {\bf 1311}, 168 (2013)
  doi:10.1007/JHEP11(2013)168
  [arXiv:1307.6498 [hep-th]].
    
\bibitem{Mamo:2015dea} 
  K.~A.~Mamo,
  JHEP {\bf 1505}, 121 (2015)
  doi:10.1007/JHEP05(2015)121
  [arXiv:1501.03262 [hep-th]].  
    
\bibitem{Dudal:2015wfn} 
  D.~Dudal, D.~R.~Granado and T.~G.~Mertens,
  Phys.\ Rev.\ D {\bf 93}, no. 12, 125004 (2016)
  doi:10.1103/PhysRevD.93.125004
  [arXiv:1511.04042 [hep-th]].
    
    \bibitem{Evans:2016jzo} 
  N.~Evans, C.~Miller and M.~Scott,
  Phys.\ Rev.\ D {\bf 94}, no. 7, 074034 (2016)
  doi:10.1103/PhysRevD.94.074034
  [arXiv:1604.06307 [hep-ph]].
    
\bibitem{Li:2016gfn} 
  D.~Li, M.~Huang, Y.~Yang and P.~H.~Yuan,
  JHEP {\bf 1702}, 030 (2017)
  doi:10.1007/JHEP02(2017)030
  [arXiv:1610.04618 [hep-th]].
    
  \bibitem{Ballon-Bayona:2017dvv} 
  A.~Ballon-Bayona, M.~Ihl, J.~P.~Shock and D.~Zoakos,
  JHEP {\bf 1710}, 038 (2017)
  doi:10.1007/JHEP10(2017)038
  [arXiv:1706.05977 [hep-th]].
    
    
\bibitem{Rodrigues:2017cha} 
  D.~M.~Rodrigues, E.~Folco Capossoli and H.~Boschi-Filho,
  arXiv:1709.09258 [hep-th].  
     
     

\bibitem{Agashe:2014kda} 
  K.~A.~Olive {\it et al.} [Particle Data Group Collaboration],
  Chin.\ Phys.\ C {\bf 38}, 090001 (2014).
  



\bibitem{DHoker:2009mmn} 
  E.~D'Hoker and P.~Kraus,
  JHEP {\bf 0910}, 088 (2009)
  doi:10.1088/1126-6708/2009/10/088
  [arXiv:0908.3875 [hep-th]].    
    
\bibitem{DHoker:2009ixq} 
  E.~D'Hoker and P.~Kraus,
  JHEP {\bf 1003}, 095 (2010)
  doi:10.1007/JHEP03(2010)095
  [arXiv:0911.4518 [hep-th]].
        
       
\bibitem{Iqbal:2008by} 
  N.~Iqbal and H.~Liu,
  Phys.\ Rev.\ D {\bf 79}, 025023 (2009)
  doi:10.1103/PhysRevD.79.025023
  [arXiv:0809.3808 [hep-th]]. 
  
\bibitem{Finazzo:2015tta} 
  S.~I.~Finazzo,
  ``Understanding strongly coupled non-Abelian plasmas using the gauge/gravity duality,'' Phd Thesis, Universidade de S\~ao Paulo, 2015,  DOI 10.11606/T.43.2015.tde-07042015-144444 .  
  
  \bibitem{Rebhan:2011vd} 
  A.~Rebhan and D.~Steineder,
  Phys.\ Rev.\ Lett.\  {\bf 108}, 021601 (2012)
  doi:10.1103/PhysRevLett.108.021601
  [arXiv:1110.6825 [hep-th]].
    
\bibitem{Adare:2014hje} 
  A.~Adare {\it et al.} [PHENIX Collaboration],
  Phys.\ Rev.\ C {\bf 91}, no. 2, 024913 (2015)
  doi:10.1103/PhysRevC.91.024913
  [arXiv:1404.2246 [nucl-ex]].
 
    
  
\bibitem{Dudal:2014jfa} 
  D.~Dudal and T.~G.~Mertens,
  Phys.\ Rev.\ D {\bf 91}, 086002 (2015)
  doi:10.1103/PhysRevD.91.086002
  [arXiv:1410.3297 [hep-th]].

\end{thebibliography}
 \end{document}